\documentclass[%
 reprint,
 amsmath,amssymb,
 aps,
]{revtex4-2}

\usepackage{graphicx}
\usepackage{dcolumn}
\usepackage{bm}
\usepackage{mathtools} 
\usepackage{amsfonts}
\usepackage{amsmath}
\usepackage{graphicx}
\usepackage{array}
\usepackage{multirow}
\usepackage{verbatim}

\newcommand{\ii}{{\rm i}}

\newcommand{\be}[0]{\begin{equation}}
\newcommand{\ee}[0]{\end{equation}}
\newcommand{\bea}[0]{\begin{eqnarray}}
\newcommand{\eea}[0]{\end{eqnarray}}

\begin{document}

\preprint{APS/123-QED}

\title{Theoretical analysis of quantum random walks with stress-engineered optics}

\author{Kevin Liang}
 \email{kliang3@ur.rochester..edu}
\author{Ashan Ariyawansa}%
\author{Thomas G. Brown}
\affiliation{The Institute of Optics, University of Rochester, 275 Hutchinson Rd, Rochester, NY 14627, USA
}%


\author{Omar S. Maga\~{n}a-Loaiza}
\affiliation{
Department of Physics and Astronomy, Louisiana State University, 202 Nicholson Hall, Baton Rouge, LA 70803, USA
}%

\date{\today}

\begin{abstract}
Quantum random walks (QRWs) are random processes in which the resulting probability density of the "walker" state, whose movement is governed by a "coin" state, is described in a non-classical manner. Previously, Q-plates have been used to demonstrate QRWs with polarization and orbital angular momentum playing the role of coin and walker states, respectively. In this theoretical analysis, we show how stress-engineered optics can be used to develop new platforms for complex QRWs through relative simple optical elements. Our work opens up new paths to speed up classical-to-quantum transitions in robust photonic networks.
\end{abstract}

\maketitle


\section{Introduction}
Random walks (RWs) are probabilistic processes where, at each discrete iteration of the walk, there is a change in the "walker" state through a change in the "coin" state. As the most basic mathematical example, the "walker" and "coin" states are literally represented by a walker and a coin: at each iteration of the RW, the coin is flipped and lands either heads or tails, which results in the walker stepping to the right or left, respectively. This procedure occurs $N$ times and the final location of the walker is expected to be given by a binomial distribution; this distribution is the hallmark of a \textit{classical} RW (CRW).

To extend this phenomenon to optical beams, one must re-imagine what the "walker" and the "coin" states are. As is apparent in the aforementioned example, the "walker" state resides in a countably infinite space and the "coin" state resides in a binary space (although it need not be). Two properties of an optical beam that fit into these criteria are orbital angular momentum (OAM) and polarization, respectively. In this work, we will label the OAM states as $e^{\ii m \phi}$, where $m \in \mathbb{Z}$, and $\phi$ is the azimuthal coordinate of a beam in cylindrical coordinates. These states correspond to the azimuthal parts of the well-known Laguerre-Gauss (LG) modes. As for polarization, we will work in the circular basis, with $(1 \, \, 0)$ and $(0 \, \, 1)$ basis vectors corresponding to left-hand circular (LHC) and right-hand circular (RHC) polarizations. Throughout this work, we will work in the language of Jones matrices (which makes clear the spin-orbit coupling of polarization and OAM) and vectors to represent optical elements and electric field, respectively.

\begin{figure}
    \centering
    \includegraphics[scale=0.5]{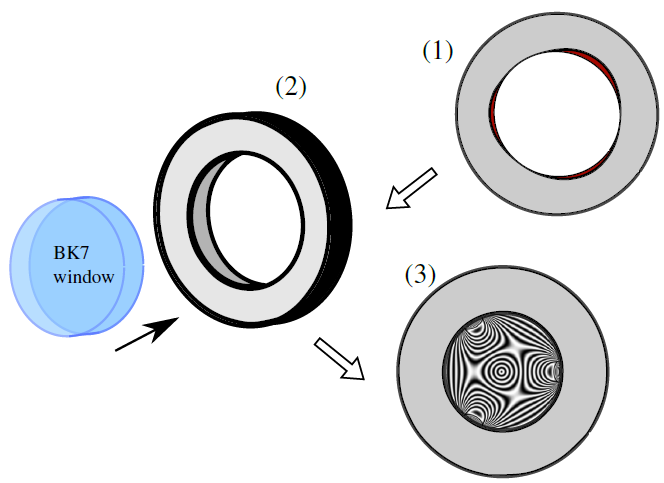}
    \caption{In a three-step process, a metal ring (with material removed symmetrically from the inner edge) is heated, BK7 glass is placed within the ring, and the metal ring is allowed to cool. The resulting compression of the BK7 from thermal cooling gives rise to a pattern of stress-induced birefringence.}
    \label{SEOfab}
\end{figure}

It has been demonstrated in the past that Q-plates and wave plates can be used to demonstrate \textit{quantum} RWs (QRWs) \cite{aharonov, sephton,cardano,travaglione,kempe,jeong,rafsanjani}. QRWs offer a robust platform to investigate dynamical processes in several fields such as chemistry, biology, sociology, and information science. It has been shown that QRWs represent a universal computational primitive with enormous potential to enable new quantum algorithms for simulation of quantum phenomena in multiple systems \cite{childs, lovett, childs2}. The key difference between CRWs and QRWs is that the various paths through the "walker" space can interfere with each other in the latter (due to the wave nature of electric fields). Hence, the respective probability densities that describe the walker's position after $N$ coin flips can be very different; those corresponding to QRWs usually display a larger variance. Immediately, it is evident that QRWs can play a role in modeling phenomena such as diffusive and ballistic transport in photonic networks. In this work, we demonstrate theoretically how a stressed engineered optic (SEO), along with wave plates, can be used to demonstrate QRWs (a diagram of an SEO is seen in Fig.~\ref{SEOfab}). 

\begin{figure}
    \centering
    \includegraphics[scale=1]{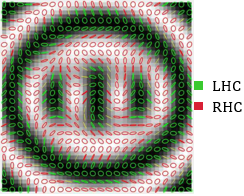}
    \caption{After a uniform LHC electric field passes through a tilted SEO, the resulting intensity of the RHC component of the output beam (shown in this figure along with spatially-varying polarization ellipses) exhibits several optical (and polarization) vortices.}
    \label{SEOexample}
\end{figure}

SEOs have been used for the past decade in various applications such as polarimetry and generation of optical beams with unconventional polarization states \cite{beckley,spilman,ram,zimmerman,sivankutty,spilman2}. Recently, they have been shown to generate beams with multiple singularities (both phase vortices and polarization C-points) \cite{ariyawansa,liang}. Figure~\ref{SEOexample} shows a result from \cite{liang}, where a uniform LHC electric field passes through a tilted SEO. The resulting RHC component of the output beam has multiple optical vortices. Although the analysis done in this work will focus on normal incidence (in which case there is only a vortex structure at the center of the output beam), Fig.~\ref{SEOexample} demonstrates the possibility of the SEO allowing for multiple QRWs simultaneously. Experimentally, there are a multitude of ways to implement QRWs depending on the degrees of freedom that compose the ``walker"-state. However, we mention here SEO-driven QRWs can be performed through stable common path interferometers.

The Jones matrix for the SEO, written in the circular basis, is given by
\begin{align}
\hat{J}_\text{SEO} = \begin{bmatrix} \cos\left[\Delta(\rho) \right] & \ii e^{\ii \phi} \sin\left[\Delta(\rho) \right] \\ \ii e^{-\ii \phi} \sin\left[\Delta(\rho) \right]& \cos \left[\Delta(\rho) \right] \end{bmatrix}, \label{jseo}
\end{align}
where $\rho$ and $\phi$ are the radial (normalized by the SEO aperture radius; so $\rho \in [0,1]$) and azimuthal coordinates, respectively. Furthermore, $\Delta(\rho) = c\rho/2$, where $c$ is the SEO's stress parameter and governs the strength of the SEO's birefringence. Note that the birefringence of the SEO depends explicitly on $\rho$. This is in contrast to the Q-plate, whose Jones matrix is given by
\begin{align}
    \hat{J}_\text{Q} = \begin{bmatrix} 0 & \ii e^{\ii 2 q \phi} \\ \ii e^{-\ii 2 q \phi} & 0 \end{bmatrix}, \label{jqplate}
\end{align}
where $q$ can be an integer or a half-integer value. 

In addition to the SEO, wave plates will also be needed to perform a QRW; the Jones matrix of a quarter-wave plate (QWP) and a half-wave plate (HWP) are given by
\begin{align}
\hat{J}_\text{QWP}(\theta) &= \frac{1}{\sqrt{2}} \begin{bmatrix} 1 & \ii e^{\ii 2\theta} \\ \ii e^{-\ii 2 \theta} & 1 \end{bmatrix}, \label{jQWP}
\end{align}
and
\begin{align}
\hat{J}_\text{HWP}(\theta) &= \ii \begin{bmatrix} 0 & e^{\ii 2\theta} \\ e^{-\ii 2\theta} & 0 \end{bmatrix}, \label{jHWP}
\end{align}
respectively, where $\theta$ is the angle between the horizontal axis and the wave plate's fast axis.

\section{Quantum random walks with the SEO}
\subsection{Theoretical proposal for generating QRWs}

\begin{figure}
    \centering
    \includegraphics[scale=.5]{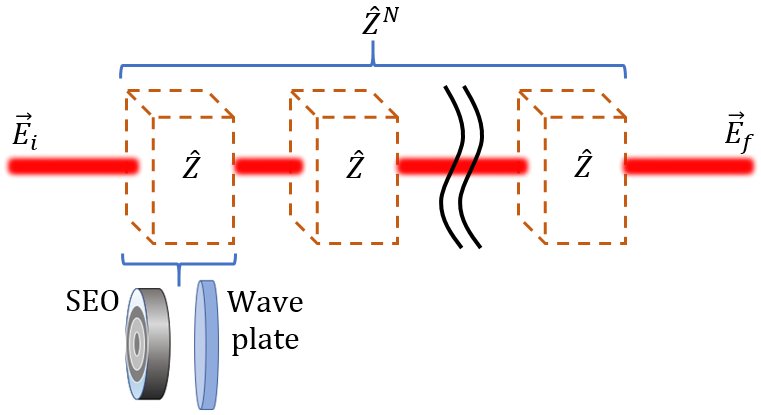}
    \caption{An illustration of the experimental implementation of QRW with SEOs. The initial field $\vec{E}_i$ goes through $N$ copies of $\hat{Z}$, which is composed of a SEO and a waveplate (either a QWP or a HWP).}
    \label{setup}
\end{figure}

In this work two types of QRWs, created by using an SEO and a wave plate, are explored. The first, called an \textit{identity} QRW, is constructed by using a HWP with $\theta = 0$:
\begin{align}
\hat{Z}_\text{I} = \hat{J}_\text{HWP}(0)  \hat{J}_\text{SEO},\label{identityQRW}
\end{align}
where $\hat{Z}_\text{I}$ is the Jones matrix that represents one step in the identity QRW. The second, called a \textit{Hadamard} QRW, is constructed by using a QWP with $\theta = \pi/4$:
\begin{align}
\hat{Z}_\text{H} = \hat{J}_\text{QWP}(\pi/4)  \hat{J}_\text{SEO},\label{hadamardQRW}
\end{align}
where $\hat{Z}_\text{H}$ is the Jones matrix that represents one step in the Hadamard QRW.

Given an initial electric field $\vec{E}_\text{i}$, the final electric field after $N$ iterations of the QRW is given by
\begin{align}
\vec{E}_\text{f}(\rho,\phi) = \hat{Z}^N \vec{E}_\text{i}(\rho,\phi),
\end{align}
where $\hat{Z}$ is either the identity or the Hadamard QRW matrix given in Eqs.~(\ref{identityQRW}) and (\ref{hadamardQRW}). A schematic for applying the $\hat{Z}^N$ on an initial beam is shown in Fig~\ref{setup}. Since we are implementing QRWs with OAM as the degree of freedom, it is beneficial to express the components of $\vec{E}_\text{f}$ as a superposition of the OAM eigenstates:
\begin{align}
\vec{E}_\text{f}(\rho,\phi) = \sum_{m=-\infty}^\infty \alpha_n(\rho) e^{\ii m \phi} \begin{bmatrix} 1 \\ 0 \end{bmatrix} + \sum_{m=-\infty}^\infty \beta_n(\rho) e^{\ii m \phi} \begin{bmatrix} 0 \\ 1\end{bmatrix}
\end{align}
where the we have explicitly written out the $\rho$-dependence of the coefficients $\alpha_n$ and $\beta_n$. This is in contrast to QRWs generated using Q-plates where these coefficient would be independent of $\rho$. Moreover, prior to OAM measurement, a polarizer will be used to allow only one (circular) component of $\vec{E}_\text{f}$ through. For instance, suppose the RHC component of $\vec{E}_\text{f}$ is blocked out. If a measurement on OAM is now performed, the probability that the OAM value of $m$, which corresponds to $e^{\ii m \phi}$, is measured is given by
\begin{align}
\mathcal{P}(m) = \frac{1}{\mu} \int_0^1 \rho \left| \alpha_m (\rho) \right|^2 \, \text{d} \rho, \label{probM}
\end{align}
the probability mass function, where 
\begin{align}
\mu = \sum_{m=-\infty}^\infty \int_0^1 \rho \left| \alpha_m (\rho) \right|^2 \, \text{d} \rho, \label{muM}
\end{align}
is a normalization factor. The analogous formulas hold true if the LHC component of $\vec{E}_\text{f}$ is blocked instead by replacing $\alpha_m$ with $\beta_m$. in Eqs.~(\ref{probM}) and (\ref{muM}).

\begin{figure}
\centering
\includegraphics[scale=0.5]{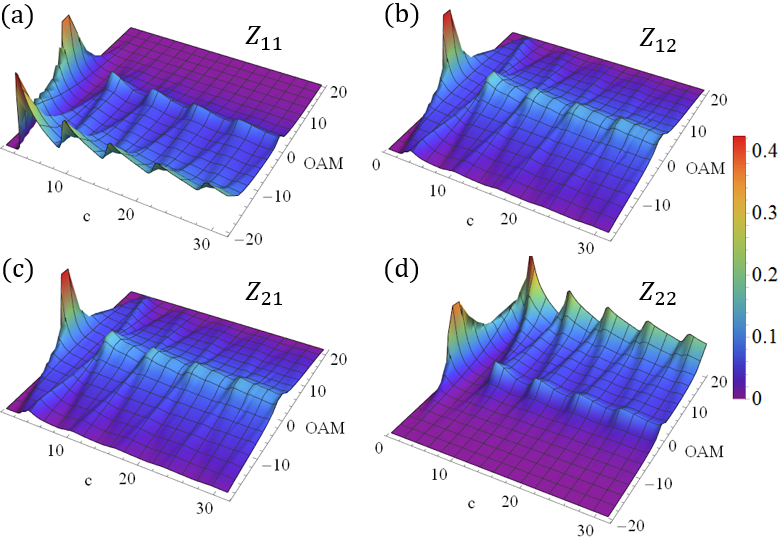}
\caption{These plots show $\int_0^1|Z_{ij}|^2 \rho\, \text{d} \rho$, over possible OAM values ($N = 20$) for various values of $c$ in an identity QRW process. Parts (a)-(d) correspond to $Z_{11}, Z_{12}, Z_{21},$ and $Z_{22}$, respectively. Note that every plot shown here approaches, in an oscillatory manner, an asymptotic value as $c$ gets large [as indicated by Eq.~(\ref{oscillatingresult})]. }
\label{highcasymptotic}
\end{figure}

Consider the explicit representation of the $N$-th power of an arbitrary QRW matrix, $\hat{Z}$:
\begin{align}
\hat{Z}^N &= \begin{bmatrix} \mathcal{Z}_{11}(N,\rho,\phi) & \mathcal{Z}_{12}(N,\rho,\phi) \\ \mathcal{Z}_{21}(N,\rho,\phi) & \mathcal{Z}_{22} (N,\rho,\phi)    \end{bmatrix}  \nonumber\\
&= \begin{bmatrix} \sum_{m} Z_{11}(N,m,\rho) e^{\ii m \phi} & \sum_m Z_{12} (N,m,\rho) e^{\ii m \phi} \\ \sum_{m} Z_{21}(N,m,\rho) e^{\ii m \phi} & \sum_m Z_{22} (N,m,\rho) e^{\ii m \phi} \end{bmatrix}, \label{arbitrarypower}
\end{align}
where $\hat{Z}$ can once again be given by either $\hat{Z}_\text{I}$ or $\hat{Z}_\text{H}$. Note that $\hat{Z}^N$, the total transition Jones matrix, represents $N$ steps through a QRW process. Given $\vec{E}_\text{i}$, it is possible to use Eq.~(\ref{arbitrarypower}) to calculate the resulting probability densities (or wavefunction amplitudes) $\alpha_m(\rho)$ and $\beta_m(\rho)$ for OAM in the components of $\vec{E}_\text{f}$. The explicit expressions for the components of $\hat{Z}^N$ are shown in Appendix~\ref{components}, and are rather unruly, so we will continue to use the concise notation presented in the second line of Eq.~(\ref{arbitrarypower}). It should be noted that $|Z_{ij}(N,m,\rho)|^2$ is periodic in $\rho$, and the number of cycles that occur within the aperture radius is given by $c/(2\pi)$, which is the same as the number of cycles of the spatial oscillations seen in matrix components of Eq.~(\ref{jseo}). Immediately, we observe that the probability densities resulting from $\hat{Z}^N$ will approach an asymptotic limit for large values of $c$ for all OAM values. This is because, for any periodic function with unit period, $f(\rho)$, we have that
\begin{align}
    \lim_{T\rightarrow 0}\int_0^1|f(\rho/T)|^2\, \text{d} \rho \rightarrow \int_0^1 |f(\rho)|^2\, \text{d} \rho. \label{oscillatingresult}
\end{align}
Equation~(\ref{oscillatingresult}) states that the average value of a periodic function over a domain of unit length, in the limit that its period approaches zero (so there are many cycles over the considered interval), is equal to the average value of the function over an interval that contains exactly one cycle. Figure~\ref{highcasymptotic} shows this behavior for the various $Z_{ij}$ coefficients, which demonstrates that the OAM probability densities generated by SEOs are insensitive to $c$, once it is sufficiently high ($c\approx 2\pi$). It should be noted that the plotted quantities are $\int_0^1 |Z_{ij}|^2 \rho \text{d} \rho$, which do not themselves correspond to OAM probability densities (since these are simply integrals over the elements of $\hat{Z}^N$), but they illustrate the global behavior of the QRW regardless of the form of $\vec{E}_i$. However, one can interpret the plots in Fig.~\ref{highcasymptotic} as $\mathcal{P}(m)$ by noting that Fig.~\ref{highcasymptotic}(a/c) give $\mathcal{P}(m)$ for the LHC/RHC components of $\vec{E}_\text{f}$ for $\vec{E}_\text{i} = (1\,\,0)$. The same is true for Fig.~\ref{highcasymptotic}(b/d) for $\vec{E}_\text{i} = (0\,\,1)$. 

\begin{figure}
    \centering
    \includegraphics[scale=0.5]{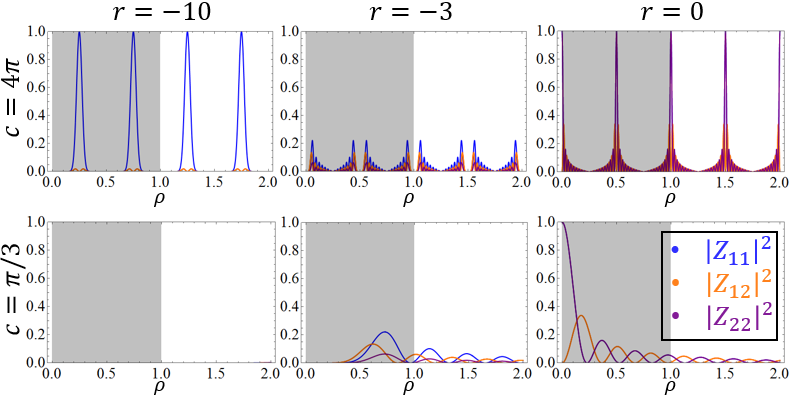}
    \caption{These plots show $|Z_{ij}|^2$ for $N=20$ and the identity QRW. Appendix~\ref{components} shows that $Z_{12} = Z_{21}$ and so only $|Z_{12}|^2$ is shown. Furthermore, the off-diagonal elements of $\hat{Z}^N$ only contain terms proportional to $e^{\ii(2r+1)\phi}$, where $r$ takes integer values in the range $[-N/2,N/2]$; that is, they only contain odd-valued OAM terms. Conversely, the diagonal elements contain terms proportional $e^{\ii 2r \phi}$. The first and second rows correspond to $c=4\pi$ and $c= \pi/3$, respectively. The gray region marks the region within the aperture. As $r$ goes from $-N/2$ to $0$, the functions spread out; this behavior is particularly relevant for the low $c$ regime. Although the range of $r \in [0,N/2]$ is not shown explicitly, the behavior of $Z_{ij}$ around $r = 0$ is symmetric, with the caveat that $Z_{11}$ and $Z_{22}$ swapping roles. }
    \label{highcvslowc}
\end{figure}

On the other hand, as can be gleaned from the low $c$ values of the plots in Fig.~\ref{highcasymptotic}, the behavior when $c$ is low (such as $c < \pi$) is quite different and worth discussing. To appreciate this phenomenon, Fig.~\ref{highcvslowc} illustrates a clear change of behavior when $c$ is low enough so that there is less than one cycle within the aperture (for $\rho \in [0,1]$). The top row of Fig.~\ref{highcvslowc} shows the case of $c = 4\pi$, in which there are two full cycles within the aperture; there, the quantity $|Z_{ij}|$ spreads out as $m$ varies from its most extreme values to the central value of $m = 0$, where there is a sudden spike at the edges of each cycle. The bottom row, on the other hand, shows the case of $c = \pi/3$. Since $2\pi/c<1$ in this case, there is not a full cycle within the aperture region. Hence, there is nearly zero probability of measuring an extreme OAM value because the $|Z_{ij}|$ functions have not spread out far enough yet to reach the aperture region. The resulting differences in the probability densities for OAM values are illustrated in Sec.~\ref{simulations}.

\subsection{Simulation results} \label{simulations}

\subsubsection{High $c$ values, and comparison to Q-plates}
\begin{figure}
\centering
\includegraphics[scale=0.38]{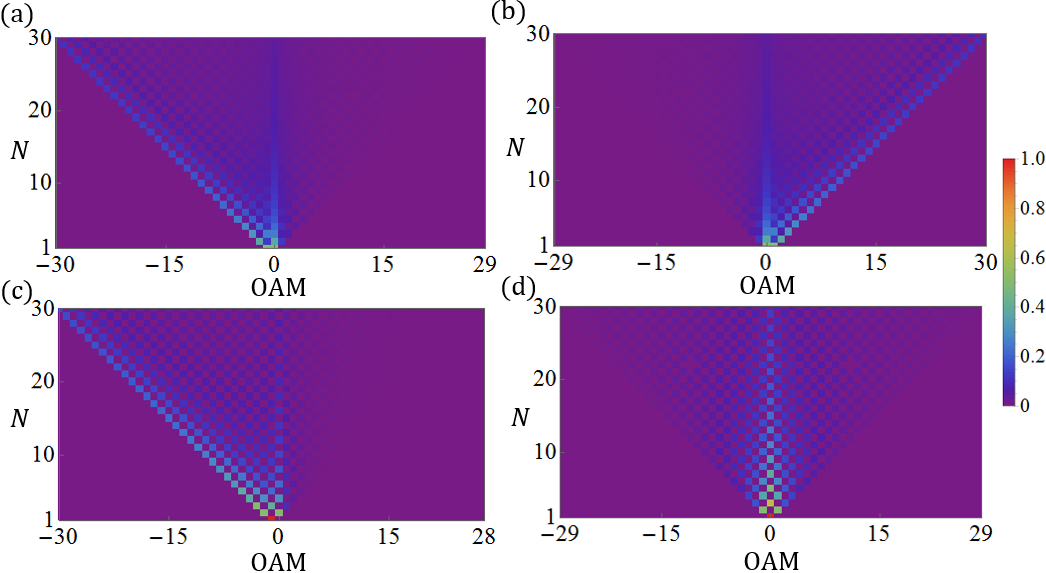}
\caption{$\mathcal{P}(m)$ for the value of OAM after $N$ iterations of the SEO identity QRW process, with $c = 4\pi$. (a) and (b) correspond to the LHC and RHC of $\vec{E}_\text{f}$ for a symmetric input field given by $\vec{E}_\text{i} = E_0 (1 \, \, 1)$. (c) and (d) correspond to the LHC and RHC of $\vec{E}_\text{f}$ for an asymmetric input field given by $\vec{E}_\text{i} = E_0 (1 \, \, 0)$. The analogous plots of a Q-plate would only have non-zero probabilities for the extreme-most values of OAM ($m = \pm N$).}
\label{identitySEOhighc}
\end{figure}

Figure~\ref{identitySEOhighc} shows the probability density plots for the identity QRW, for various input electric fields with $c = 4\pi$. One can observe from Figs.~\ref{identitySEOhighc}(a) and (b) that an initial uniform electric field $\vec{E}_\text{i}$ with equal amounts of LHC and RHC components yields a $\vec{E}_\text{f}$ whose components each give rise to an asymmetric (in OAM) probability distribution in OAM. On the otherhand, an initial uniform electric field with only a LHC component gives rise a different set of probability densities in the components of the output field; this is seen in Figs.~\ref{identitySEOhighc}(c) and (d).

It should be noted that, for Q-plates, the identity QRW gives rise to the QRW with the greatest variance \cite{sephton}. That is, $N$ increased, the only OAM values that have non-zero probabilities were those that were the furthest from $l = 0$. This is in contrast with the SEO result, as seen in Fig.~\ref{identitySEOhighc}. There, it is apparent that there is a continuum of possible measurable OAM states after $N$ iterations. For instance, Fig.~\ref{identitySEOhighc}(a) shows that the LHC of $\vec{E}_\text{f}$, for a symmetric $\vec{E}_\text{i}$, has probabilities that peak for extreme negative values of $l$ and $l = 0$. These results show a marked difference between the QRWs generated by Q-plates and SEOs.

\begin{figure}
\centering
\includegraphics[scale=0.38]{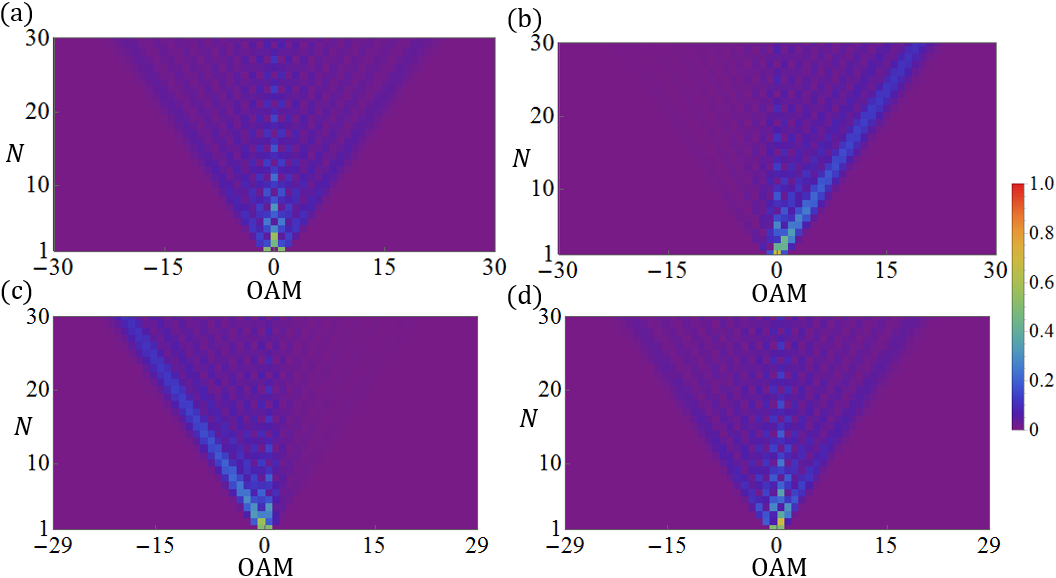}
\caption{$\mathcal{P}(m)$ for the value of OAM after $N$ iterations of the SEO Hadamard QRW process, with $c = 4\pi$. (a) and (b) correspond to the LHC and RHC of $\vec{E}_\text{f}$ for a symmetric input field given by $\vec{E}_\text{i} = E_0 (1 \, \, 1)$. (c) and (d) correspond to the LHC and RHC of $\vec{E}_\text{f}$ for an asymmetric input field given by $\vec{E}_\text{i} = E_0 (1 \, \, 0)$. }
\label{hadamardSEOhighc}
\end{figure}

\begin{figure}
    \centering
    \includegraphics[scale=0.38]{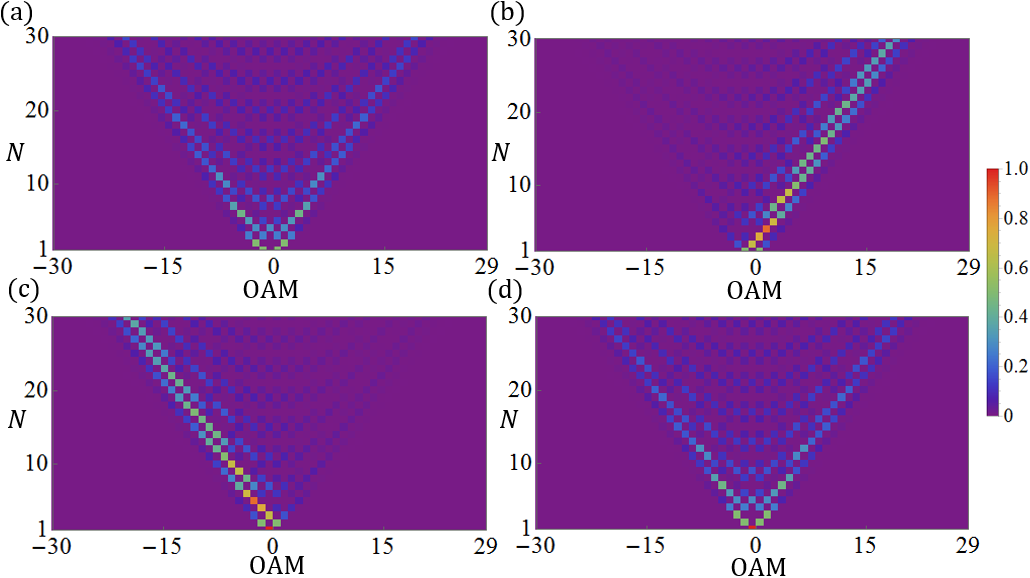}
    \caption{$\mathcal{P}(m)$ for the value of OAM after $N$ iterations of the Hadamard QRW process, for a Q-plate with $q = 1/2$. (a) and (b) correspond to the LHC and RHC of $\vec{E}_\text{f}$ for a symmetric input field given by $\vec{E}_\text{i} = E_0 (1 \, \, 1)$. (c) and (d) correspond to the LHC and RHC of $\vec{E}_\text{f}$ for an asymmetric input field given by $\vec{E}_\text{i} = E_0 (1 \, \, 0)$. }
    \label{hadamardqplate}
\end{figure}

Figure~\ref{hadamardSEOhighc} shows the probability mass plots for the Hadamard QRW, for various input electric fields with $c =4\pi$. 
Like the Q-plates, the Hadamard QRW with SEOs give rise to OAM probability distributions that populate many values of $l$. However, it should be noted that these probability distributions are very different in appearance, as we can see by comparing Fig.~\ref{hadamardSEOhighc} with Fig.~\ref{hadamardqplate}, which shows the analogous plots, but for a Q-plate with $q = 1/2$. It is evident that both the SEO and Q-plate illustrate the same types of symmetry properties in the output beam, but the qualitative behavior is markedly different. 

\begin{figure}
    \centering
    \includegraphics[scale=0.7]{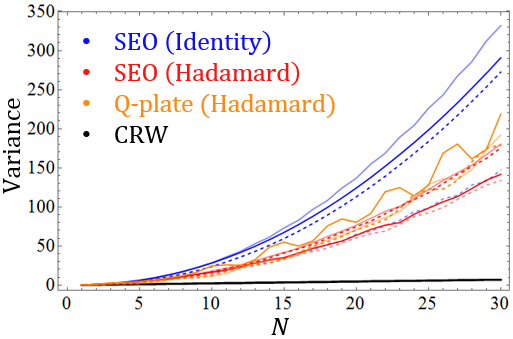}
    \caption{The variances, as a function of $N$, of the probability mass function $\mathcal{P}(m)$ associated with various RW processes. For the QRWs, the opaque and translucent curves correspond to the LHC and RHC components of the output beam, respectively. Furthermore, the solid and dashed curves correspond to $\vec{E}_\text{i} = E_0 (1\,\, 1)$ and $E_0(1\,\,0)$, respectively. It should be noted that, although not shown here, the identity QRW with the Q-plate gives the largest variances.}
    \label{variances}
\end{figure}

Finally, to emphasize the difference between the QRWs produced by SEOs (and Q-plates) and CRWs, we compare the variances of the corresponding probability mass functions $\mathcal{P}(m)$. Figure~\ref{variances} illustrates that the variances associated with QRWs grow much quicker than that of the CRW.

\subsubsection{Low $c$ values}

\begin{figure}
\centering
\includegraphics[scale=0.38]{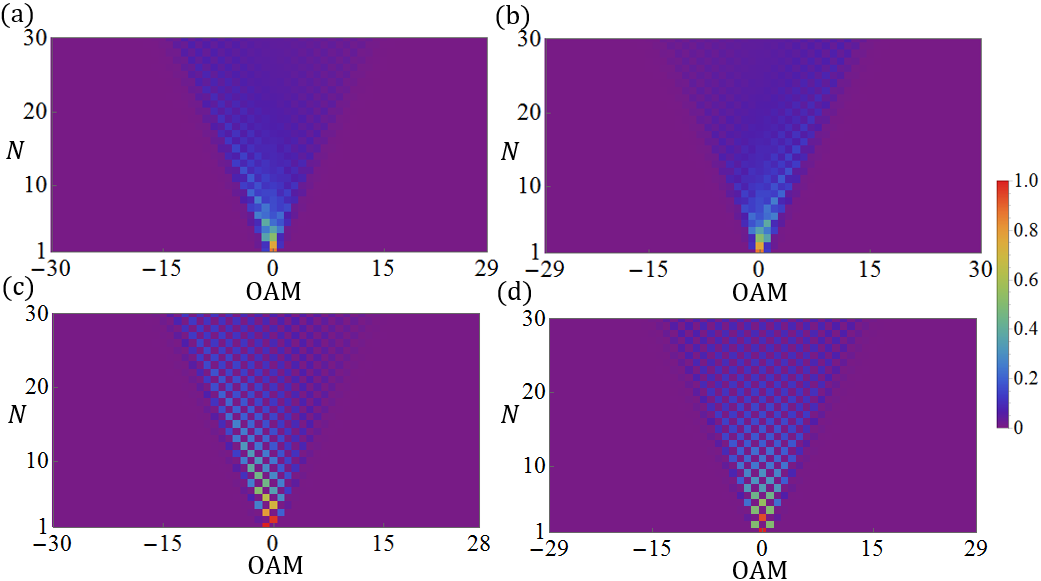}
\caption{$\mathcal{P}(m)$ for the value of OAM after $N$ iterations of the SEO identity QRW process, with $c = \pi/3$. (a) and (b) correspond to the LHC and RHC of $\vec{E}_\text{f}$ for a symmetric input field given by $\vec{E}_\text{i} = E_0 (1 \, \, 1)$. (c) and (d) correspond to the LHC and RHC of $\vec{E}_\text{f}$ for an asymmetric input field given by $\vec{E}_\text{i} = E_0 (1 \, \, 0)$. }
\label{identitySEOlowc}
\end{figure}

To illustrate how $\mathcal{P}(m)$ can exhibit different behaviors for low values of $c$, we perform simulations akin to those seen in Fig.~\ref{identitySEOhighc} (the identity QRW) but with $c= \pi/3$. Figure~\ref{identitySEOlowc} shows these results. A notable difference between the plots in Fig.~\ref{identitySEOhighc} and Fig.~\ref{identitySEOlowc} concerns the range of the characteristic fan-shape that indicates the possible spread of OAM values: this spread is smaller for the lower $c$ value. Moreover, although the discussions regarding asymmetric and symmetry are still true, the peak at $m  = 0$ is no longer visible in Fig.~\ref{identitySEOlowc} as compared to Fig.~\ref{identitySEOhighc}. 

\begin{figure}
\centering
\includegraphics[scale=0.38]{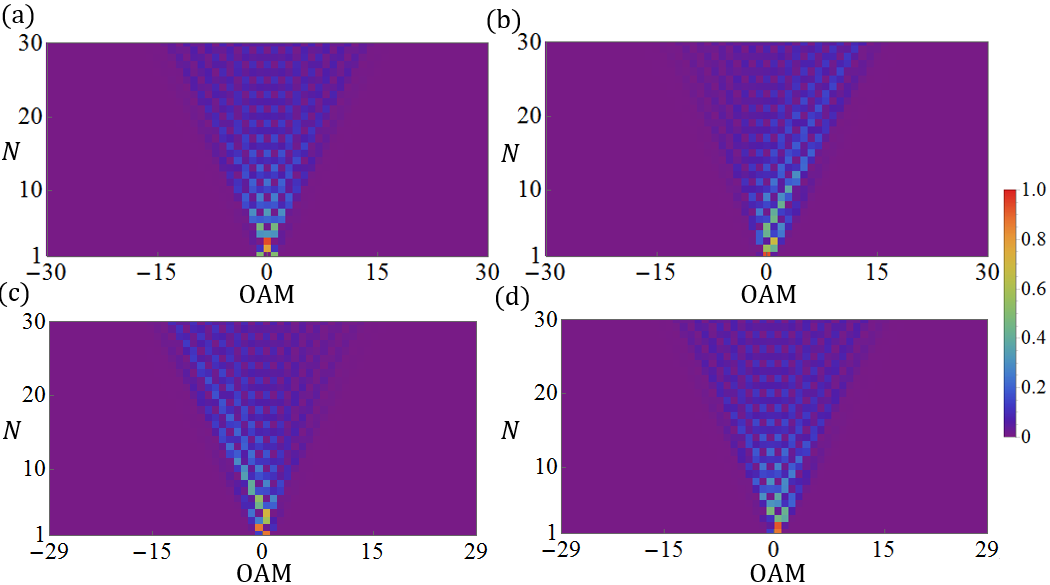}
\caption{$\mathcal{P}(m)$ for the value of OAM after $N$ iterations of the SEO Hadamard QRW process, with $c = \pi/3$. (a) and (b) correspond to the LHC and RHC of $\vec{E}_\text{f}$ for a symmetric input field given by $\vec{E}_\text{i} = E_0 (1 \, \, 1)$. (c) and (d) correspond to the LHC and RHC of $\vec{E}_\text{f}$ for an asymmetric input field given by $\vec{E}_\text{i} = E_0 (1 \, \, 0)$. }
\label{hadamardSEOlowc}
\end{figure}

For completeness, we also show the simulations for $c = \pi/3$ in a Hadamard QRW process. These are shown in Fig.~\ref{hadamardSEOlowc}. Once again, we see that the spread of $\mathcal{P}(m)$ is smaller as compared to Fig.~\ref{hadamardSEOhighc} and the prominent peak at $m = 0$ is missing for smaller values of the SEO's stress parameter. Furthermore, the symmetry and asymmetry discussions that were mentioned earlier for Fig.~\ref{hadamardSEOhighc} are still valid here.

\section{Concluding remarks and discussion} \label{concludingremarks}
We have shown how SEOs may be used to generate QRWs in a similar fashion to those that have been done with Q-plates in the past. Although the processes, and possibly experimental procedures, are very similar to QRWs involving Q-plates, there are several novel aspects that result directly from the radial dependence of the SEO's spatial birefringence. Because of this, the probability distributions are more complicated to calculate and new aspects, such as the behavior of $\mathcal{P}(m)$ in the high and low $c$ regimes, are revealed. Although the simulations shown here were done specifically for the identity and Hadamard QRW processes, a more general analysis can be done using the formalism presented in \cite{sephton}. In essence, the analysis done here serves to expand the realm of possibilities for what can be done with the OAM degree of freedom in the context of QRWs. While Q-plates provide a simple example of spin-orbit coupling that can give rise to QRWs, we have shown here that SEOs, which are easily manufactured, can demonstrate the same phenomena.

In addition to pointing out the similarities to Q-plates, it should also be noted that the Jones matrix for the SEO has non-zero diagonal elements, which is in contrast to Q-plates. This allows from the analysis of multi-path QRWs, since a purely circularly-polarized input beam will acquire OAM content in both components of the output beam. Furthermore, it has been shown, for instance in Fig.~\ref{SEOexample}, that oblique beam propagation through an SEO results in multiple polarization (and phase) singularities of the output beam \cite{liang}. Hence, one can envision performing multiple QRWs in parallel, one for each of the many phase vortices created through oblique propagation. This opens the path for new theoretical and experimental considerations that involve not only SEOs in normal incidence (considered in this work), but also of those in which the electric field passes through obliquely. The field angle of incidence through an SEO plays an important role in the vortex structure of the output beam because the SEO is a thick optical element, which is a property that the Q-plate does not possess.

\appendix
\section{Expressions for the matrix elements of the total transition Jones matrix} \label{components}
In this appendix, we will calculate explicit expressions for the matrix components of $\hat{Z}^N$, as seen in Eq.~(\ref{arbitrarypower}). We will consider two special cases where the waveplate element in the QRW process is either a HWP or a QWP, with arbitrary fast-axis orientation angle $\theta$.

\subsection{QRWs with a HWP}
For a QRW construction that involves a HWP (with arbitrary fast axis angle), $\hat{Z} = \hat{J}_{\text{HWP}}(\theta) \hat{J}_\text{SEO}$.
By finding the eigenvalues $\lambda_1$ and $\lambda_2$ of $\hat{Z}$ and using the two-dimensional matrix result of
\begin{align}
\hat{Z}^N = \frac{\lambda_2 \lambda_1^N - \lambda_1 \lambda_2^N}{\lambda_2 - \lambda_1} \hat{I} + \frac{\lambda_2^N - \lambda_1^N}{\lambda_2 - \lambda_1} \hat{Z}, \label{2by2}
\end{align}
where $\hat{I}$ is the identity matrix, we find the following expressions for the matrix elements for $\hat{Z}^N$:
\begin{align}
&\mathcal{Z}_{11}(N,\rho,\phi)= \nonumber\\
& \sum_r \left\{ \mathcal{A}_{N,r}(\rho,\theta) + e^{\ii 2\theta} \sin[\Delta(\rho)] \mathcal{A}_{N+1,r+1/2}(\rho,\theta) \right\} e^{\ii 2r \phi}, \label{Z11hwp} \\
&\mathcal{Z}_{12}(N,\rho,\phi)= \nonumber\\
& \sum_r \left\{- \ii e^{\ii 2\theta} \cos[\Delta(\rho)] \mathcal{A}_{N+1,r + 1/2}(\rho,\theta) \right\} e^{\ii (2r+1)\phi}, \label{Z12hwp}\\
&\mathcal{Z}_{21}(N,\rho,\phi)= \nonumber\\
& \sum_r\left\{- \ii e^{-\ii 2\theta} \cos[\Delta(\rho)] \mathcal{A}_{N+1,r + 1/2}(\rho,\theta) \right\} e^{\ii (2r+1)\phi}, \label{Z21hwp}\\
&\mathcal{Z}_{22}(N,\rho,\phi)= \nonumber\\
& \sum_r \left\{ A_{N,r}(\rho,\theta) + e^{-\ii 2\theta} \sin[\Delta(\rho)] \mathcal{A}_{N+1,r-1/2}(\rho,\theta) \right\} e^{\ii 2r\phi}. \label{Z22hwp}
\end{align}
Note that Eqs.~(\ref{Z11hwp})-(\ref{Z22hwp}) are written as superpositions of OAM eigenstates and the summations run from $r = -N/2$ to $r = N/2$. To understand what $\mathcal{A}_{N,r}(\rho)$ is, we first define
\begin{align}
A_{N,l,q,p}(\rho,\theta) &= \left[ \binom{n}{2l+1} - \binom{n}{2l} \right] \binom{l}{q} \binom{n-2q}{p} \nonumber\\
&\times \frac{(-1)^{n+q+1}}{2^{n-2q}}\sin^{n-2q}[\Delta(\rho)]e^{\ii 2[n-2(q+p)] \theta}. 
\end{align}
Then, we have that
\begin{align}
\mathcal{A}_{N,r}(\rho,\theta) = \sum_{s=0}^{N/2} \sum_{j=0}^{\text{Min}\left[ s, \frac{N - (|2r|-1)}{2} \right]} A_{N,s,j,r+N/2-j}(\rho).
\end{align}

\subsection{QRWs with a QWP}
For a QRW construction that involves a QWP (with arbitrary fast axis angle), $\hat{Z} = \hat{J}_\text{QWP}(\theta) \hat{J}_\text{SEO}$. 
Equation~(\ref{2by2}) can be used once again to find the matrix elements of $\hat{Z}^{N}$:
\begin{align}
&\mathcal{Z}_{11}(N,\rho,\phi)= \sum_r \bigg( \mathcal{B}_{N,r}(\rho,\theta) \nonumber\\
 &  + \frac{e^{\ii r \phi}}{\sqrt{2}} \left\{ \cos [\Delta(\rho)] \mathcal{D}_{N,r} - e^{\ii 2 \theta} \sin [ \Delta(\rho)] \mathcal{D}_{N,r+1} \right\} \bigg) , \\
&\mathcal{Z}_{12}(N,\rho,\phi)  \nonumber\\
&= \sum_r \frac{e^{\ii r \phi}}{\sqrt{2}} \left\{ \ii e^{\ii 2\theta} \cos[\Delta(\rho)] \mathcal{D}_{N,r} + \ii \sin[\Delta(\rho)] \mathcal{D}_{N,r-1} \right\} ,\\
&\mathcal{Z}_{21}(N,\rho,\phi) \nonumber\\
&= \sum_r \frac{e^{\ii r \phi}}{\sqrt{2}} \left\{ \ii e^{-\ii 2\theta} \cos[\Delta(\rho)] \mathcal{D}_{N,r} + \ii \sin[\Delta(\rho)] \mathcal{D}_{N,r+1} \right\} ,\\
&\mathcal{Z}_{22}(N,\rho,\phi)= \sum_r \bigg( \mathcal{B}_{N,r}(\rho,\theta)\nonumber\\
&   + \frac{e^{\ii r \phi}}{\sqrt{2}} \left\{ \cos [\Delta(\rho)] \mathcal{D}_{N,r} - e^{-\ii 2 \theta} \sin [ \Delta(\rho)] \mathcal{D}_{N,r-1} \right\} \bigg) .
\end{align}
Here, the summations run from $r = -N$ to $r = N$. To understand what $\mathcal{B}_{N,r}$ and $\mathcal{D}_{N,r}$ are, we first define
\begin{align}
&B_{N,l,q,p,y}(\rho,\theta) \nonumber\\
&= \left[ \binom{n}{2l+1} - \binom{n}{2l} \right] \binom{l}{q} \binom{n-2q}{p} \binom{p}{y} \frac{ (-1)^{2n-q-p+1}}{\sqrt{2}^{n-2q+2p}} \nonumber\\
&\times  \cos^{n-2q-p}[\Delta(\rho)] \sin^p[\Delta(\rho)] e^{-\ii 2 (p-2y)\theta}, \\
&D_{N,l,q,p,y}(\rho,\theta) \nonumber\\
&= \binom{n}{2l+1} \binom{l}{q} \binom{n-(2q+1)}{p} \binom{p}{y}\frac{(-1)^{2n-q-p} }{\sqrt{2}^{n-(2q+1)+2p}} \nonumber\\
&\times\cos^{n-(2q+1)-p}[\Delta(\rho)] \sin^p [\Delta(\rho)]  e^{-\ii 2 (p-2y)\theta}.
\end{align}
Then, we have that 
\begin{align}
\mathcal{B}_{N,r}(\rho,\theta) &= \sum_{l=0}^{N/2} \sum_{q=0}^l \sum_{p=0}^{n-2q} \sum_{y=0}^p B_{N,l,q,p,y}(\rho,\theta) \delta^{p-2y}_r,\\
\mathcal{D}_{N,r}(\rho,\theta) &= \sum_{l=0}^{N/2} \sum_{q=0}^l \sum_{p=0}^{N-(2q+1)} \sum_{y=0}^p D_{N,l,q,p,y}(\rho,\theta) \delta^{p-2y}_r,
\end{align}
where $\delta^a_b$ is the Kronecker delta.

\section*{Funding}
National Science Foundation (PHY-1507278).

\section*{Disclosures}
The authors declare no conflicts of interest.


\begin{thebibliography}{1}

\bibitem{aharonov}
Y. Aharonov, L. Davidovich, and N. Zagury.
\newblock ``Quantum random walks,"
\newblock \textit{Phys. Rev. A} \textbf{48}, 1687 (1993).


\bibitem{sephton}
B. Sephton, A. Dudley, G. Ruffato, F. Romanato, L. Marrucci, M. Padgett, S. Goyal, F. Roux, T. Konrad, A. Forbes.
\newblock ``A versatile quantum walk resonator with bright classical light,"
\newblock \textit{arXiv preprint arXiv:1810.06850}, (2018).

\bibitem{cardano}
F. Cardano, Francesco Massa, H. Qassim, E. Karimi, S. Slussarkenko, D. Paparo, C. de Lisio, F. Sciarrino, E. Santamato, R. W. Boyd, and L. Marrucci.
\newblock ''Quantum walks and wavepacket dynamics on a lattice with twisted photons,"
\newblock \textit{Science Advances} \textbf{1}, 2 (2015).

\bibitem{travaglione}
B. C. Travaglione and G. J. Milburn.
\newblock ``Implementing the quantum random walk,"
\newblock \textit{Phys. Rev. A} \textbf{65}, 032310 (2002).

\bibitem{kempe}
J. Kempe.
\newblock ``Quantum random walks - an introductory overview,"
\newblock \textit{arXiv preprint arXiv:quant-ph/0303081}, (2008).

\bibitem{jeong}
H. Jeong, M. Paternostro, and M. S. Kim.
\newblock ``Simulation of quantum random walks using the interference of a classical field,"
\newblock \textit{Phys. Rev. A} \textbf{69}, 012310 (2004).

\bibitem{rafsanjani}
S. Mohammad Hashemi Rafsanjani, M. Mirhosseini, O. S. Maga\~{n}a-Loaiza, and R. W. Boyd.
\newblock ``State transfer based on classical nonseparability,"
\newblock \textit{Phys. Rev. A} \textbf{92} 023827 (2015).


\bibitem{childs}
A. M. Childs.
\newblock "Universal computation by quantum walk,"
\newblock \textit{Phys. Rev. Lett.} \textbf{102} 180501 (2009).

\bibitem{lovett}
N. B. Lovett, S. Cooper, M. Everitt, M. Trevers, and V. Kendon.
\newblock "Universal quantum computation using the discrete-time quantum walk,"
\newblock \textit{Phys. Rev. A} \textbf{81} 042330 (2010).

\bibitem{childs2}
A. M. Childs, D. Gosset, and Z. Webb.
\newblock "Universal computation by multiparticle quantum walk,"
\newblock \textit{Science} \textbf{339} 6121 (2013).





\bibitem{beckley}
A. M. Beckley, T. G. Brown, and M. A. Alonso, “Full poincaré beams,” \textit{Opt. Express} \textbf{18}, 10777–10785 (2010).

\bibitem{spilman}
A. K. Spilman and T. G. Brown, “Stress birefringent, space-variant wave plates for vortex illumination,” \textit{Appl. Opt.} \textbf{46}, 61–66 (2007).

\bibitem{ram}
R. D. Ramkhalawon, T. G. Brown, and M. A. Alonso, “Imaging the polarization of a light field,” \textit{Opt. Express} \textbf{21}, 4106–4115 (2013).

\bibitem{zimmerman}
B. G. Zimmerman, R. Ramkhalawon, M. Alonso, and T. G. Brown,“Pinhole array implementation of star test polarimetry,” Proceedings of SPIE. \textbf{8949} (2014).

\bibitem{sivankutty}
S. Sivankutty,  E. R. Andresen,  G. Bouwmans,  T. G. Brown,  M. A.Alonso, and H. Rigneault, “Single-shot polarimetry imaging of multicore fiber,” \textit{Opt. Lett.} \textbf{41}, 2105–2108 (2016).

\bibitem{spilman2}
A. K. Spilman and T. G. Brown, “Stress-induced focal splitting,” \textit{Opt.Express} \textbf{15}, 8411–8421 (2007).

\bibitem{ariyawansa}
A. Ariyawansa and T. G. Brown, “Oblique propagation of light through a thick, space-variant birefringent element,” \textit{Opt. Express} \textbf{26}, 18832–18841 (2018).

\bibitem{liang}
A. Ariyawansa, K. Liang, and T. G. Brown, "Polarization singularities in a stress-engineered optic," \textit{J. Opt. Soc. Am. A} \textbf{36}, 312-319 (2019).


\end{thebibliography}
\end{document}